\documentclass[english]{llncs}
\usepackage[T1]{fontenc}
\usepackage[latin9]{inputenc}
\usepackage{color}
\usepackage{amsmath}
\usepackage{amssymb}
\usepackage{graphicx}

\makeatletter
\usepackage{algorithmicx}
\usepackage{algorithm}
\usepackage{algpseudocode}

\author{Rui Zhou \and Junyan Liu \and Sandeep Kumar \and Daniel P. Palomar}

\institute{Department of Electronic and Computer Engineering \\
The Hong Kong University of Science and Technology, Hong Kong \\
\email{\{rui.zhou, jliubl\}@connect.ust.hk, \{eesandeep, palomar\}@ust.hk}}

\makeatother

\usepackage{babel}
\begin{document}
\title{Robust Factor Analysis Parameter Estimation\thanks{This work was supported by the Hong Kong RGC 16208917 research grant.}}
\maketitle
\begin{abstract}
\vspace{-15bp}
This paper considers the problem of robustly estimating the parameters
of a heavy-tailed multivariate distribution when the covariance matrix
is known to have the structure of a low-rank matrix plus a diagonal
matrix as considered in factor analysis (FA). By assuming the observed
data to follow the multivariate Student's $t$ distribution, we can
robustly estimate the parameters via maximum likelihood estimation
(MLE). However, the MLE of parameters becomes an intractable problem
when the multivariate Student's $t$ distribution and the FA structure
are both introduced. In this paper, we propose an algorithm based
on the generalized expectation maximization (GEM) method to obtain
estimators. The robustness of our proposed method is further enhanced
to cope with missing values. Finally, we show the performance of our
proposed algorithm using both synthetic data and real financial data.

\vspace{-8bp}
\end{abstract}

\section{Introduction\vspace{-10bp}
}

Factor analysis (FA) is of great significance in various fields like
finance, statistics, and cognitive ratio \cite{sardarabadi2018complex,tsay2005analysis}.
A basic FA model can be written as $\mathbf{x}=\boldsymbol{\mu}+\mathbf{B}\mathbf{f}+\boldsymbol{\varepsilon}$,
where $\mathbf{x}\in\mathbb{R}^{p}$ is the observed vector, $\boldsymbol{\mu}\in\mathbb{R}^{p}$
is a constant vector, $\mathbf{B}\in\mathbb{R}^{p\times r}\left(r\ll p\right)$
is the factors loading matrix, $\mathbf{f}\in\mathbb{R}^{r}$ is a
vector of low-dimensional common factors, and $\boldsymbol{\varepsilon}\in\mathbb{R}^{p}$
is a vector of uncorrelated noise. For example, in a financial market,
$\mathbf{x}$ can be the return of stocks, and $\mathbf{f}$ can be
some macroeconomic factors like growth rate of the GDP, inflation
rate, unemployment rate, etc. \cite{feng2016book}.  FA model typically
assumes that $\mathbf{f}$ and $\boldsymbol{\varepsilon}$ are uncorrelated
and both zero-mean, and the covariance matrix of $\mathbf{f}$ is
an $r\times r$ identity matrix, denoted by $\mathbf{I}_{r}$. Following
this, the covariance matrix of $\mathbf{x}$ can be expressed as $\mathbf{\Sigma}=\mathbf{B}\mathbf{B}^{T}+\mathbf{\Psi}$,
where $\mathbf{\Psi}$ is a $p\times p$ diagonal matrix containing
the variance of noise at its diagonal. Note that, with the FA structure,
the number of parameters of the covariance matrix has been greatly
reduced from $p\left(p+1\right)/2$ to $p\left(r+1\right)$. Therefore,
the estimation of $\mathbf{\Sigma}$ could be improved due to the
FA structure.\vspace{-10bp}

\subsection{Related Works\vspace{-6bp}
}

\subsubsection{Learn From $\mathbf{\Sigma}$:}

A large amount of literature has focused on estimating the covariance
matrix with FA structure. One choice is decomposing the estimated
$\mathbf{\Sigma}$ into the sum of a low-rank matrix and a diagonal
matrix, e.g., constrained minimum trace factor analysis (MTFA) \cite{ten1981MTFA}.
But a main drawback is imposing the hard equality constraint $\mathbf{\Sigma}=\mathbf{B}\mathbf{B}^{T}+\mathbf{\Psi}$
, which does not allow any differences between $\mathbf{\Sigma}$
and $\mathbf{B}\mathbf{B}^{T}+\mathbf{\Psi}$. Its application is
restricted as the exact $\mathbf{\Sigma}$ is usually not observed.
Another choice is to approximate the target matrix by a FA structured
one. A naive method is to obtain $\hat{\mathbf{B}}$ firstly via
principal component analysis (PCA) and then $\hat{\mathbf{\Psi}}$
by taking directly the residual's sample variance. A joint estimation
method over $\mathbf{B}$ and $\mathbf{\Psi}$ is usually chosen to
minimize $\lVert\mathbf{\Sigma}-\left(\mathbf{B}\mathbf{B}^{T}+\mathbf{\Psi}\right)\rVert_{F}^{2}$,
where $\lVert\cdot\rVert_{F}$ denotes the Frobenius norm of a matrix.
\textcolor{black}{This problem can be solved by applying PCA iteratively
\cite{sardarabadi2018complex}.}

\vspace{-7bp}

\subsubsection{Learn From Data:}

Different from works mentioned above, the MLE for FA directly learns
the parameters from raw data. It assumes that the data are generated
from a certain statistical model, typically the multivariate Gaussian
distribution, and then the parameters are estimated by maximizing
the likelihood function. However, a disadvantage of the estimators
under the Gaussian assumption is sensitiveness to outliers \cite{feng2016book}.
A popular way to obtain a more robust estimation result is to consider
some robust heavy-tailed distribution, such as multivariate Student's
$t$ or multivariate Skew $t$ \cite{wang2017skewtFA,zhang2014tFA}
instead of Gaussian. The two aforementioned methods both assume that
$\mathbf{f}$ and $\boldsymbol{\varepsilon}$ follow the same heavy
tail distribution sharing the same degrees of freedom. As $\mathbf{f}$
and $\boldsymbol{\varepsilon}$ are not observed, such an assumption
is very restrictive and difficult to verify in practice.\vspace{-8bp}

\subsection{Contributions\vspace{-8bp}
}

 This paper considers more general and practically verifiable assumptions
on the FA model: we only assume that the observation $\mathbf{x}$
follows a multivariate Student's $t$ distribution and the FA structure
with no additional restrictions on $\mathbf{f}$ and $\boldsymbol{\varepsilon}$.
For this more general model, we propose an efficient algorithm to
estimate the parameters based on the generalized expectation maximization
(GEM) \cite{dempster1977GEM} method.   In addition, we use the
PX-EM \cite{liu1998parameter} method to accelerate the GEM. Our proposed
algorithm can be easily extended to other situations, e.g., when observed
data contains missing values \cite{liu1995ml} or when it follows
the multivariate Skew $t$ distribution.  With synthetic data, our
proposed algorithm shows great estimation accuracy and robustness
to outliers and missing data, which is very meaningful in practical
applications. We also consider real market data in the numerical results,
where the global minimum variance portfolio is designed using our
estimator and compared with those using other estimators.

\vspace{-10bp}

\section{Gaussian FA Problem\label{sec: Gaussian-FA-Problem}\vspace{-10bp}
}

Given the sample covariance matrix of the observed data as $\mathbf{S}$,
the MLE problem for FA under the Gaussian distribution assumption
(GFA) is given as below:

\vspace{-6bp}

\begin{equation}
\begin{aligned}\underset{\mathbf{\Sigma},\mathbf{B},\mathbf{\Psi}}{\mathsf{maximize}}\,\, & \quad\log\lvert\mathbf{\Sigma}^{-1}\rvert-\textrm{Tr}\left(\mathbf{\Sigma}^{-1}\mathbf{S}\right)\\
\mathsf{subject\,\,to} & \quad\mathbf{\Sigma}=\mathbf{B}\mathbf{B}^{T}+\mathbf{\Psi}\\
 & \quad\mathbf{\Psi}=\textrm{Diag}\left(\psi_{1},\dots,\psi_{p}\right)\succ\mathbf{0}.
\end{aligned}
\label{eq: Gaussian ML factor model}
\end{equation}
The solution to problem (\ref{eq: Gaussian ML factor model}) would
be $\mathbf{\Sigma}^{\star}=\mathbf{S}$ if the structure constraints
were ignored, but becomes intractable when the FA structure is imposed.
Here we introduce two algorithms for solving problem (\ref{eq: Gaussian ML factor model}).\vspace{-10bp}

\subsubsection{Alternating Algorithm:}

Problem (\ref{eq: Gaussian ML factor model}) can be solved by an
alternating optimization approach, which can be performed by alternately
updating $\mathbf{B}$ and $\mathbf{\Psi}$ (note that this actually
corresponds to an alternating optimization over a factor model decomposition
of $\mathbf{\Sigma}^{-1}$ instead of an alternating optimization
 over $\mathbf{B}$ and $\mathbf{\Psi}$, cf. \cite{David2018alternativeFA,Santamaria2017}).
For fixed $\mathbf{\Psi}$, the optimal update for $\mathbf{B}$ is
given next. \vspace{-6bp}

\begin{lemma}

\label{lem: optimal B given Psi} (\cite{David2018alternativeFA})
Given a feasible $\mathbf{\Psi}$, the optimal $\mathbf{B}^{\star}$
maximizing problem (\ref{eq: Gaussian ML factor model}) is $\mathbf{B}^{\star}=\mathbf{\Psi}^{\frac{1}{2}}\mathbf{U}\mathbf{D}^{\frac{1}{2}}$
where $\mathbf{U}\mathbf{\Lambda}\mathbf{U}^{T}$ is the eigenvalue
decomposition (EVD) of $\mathbf{\Psi}^{-\frac{1}{2}}\mathbf{S}\mathbf{\Psi}^{-\frac{1}{2}}$
and $\mathbf{D}=\textrm{Diag}\left(d_{1},\dots,d_{r},0,\dots,0\right)$
with $d_{i}=\max\left(\lambda_{i}-1,0\right)$.

\end{lemma}\vspace{-8bp}

\noindent For fixed $\mathbf{B}$, the update for $\mathbf{\Psi}$
is set as $\textrm{Diag}\left(\mathbf{S}-\mathbf{B}\mathbf{B}^{T}\right)$.
\vspace{-10bp}

\subsubsection{MM Algorithm:}

Recently, a majorization-minimization (MM) \cite{sun2017majorization}
based method has been proposed in \cite{khamaru2018computation} to
obtain the optimal $\mathbf{\Sigma}$. Plugging the optimal $\mathbf{B}^{\star}$
from Lemma \ref{lem: optimal B given Psi} in (\ref{eq: Gaussian ML factor model}),
we can achieve a concentrated version of (\ref{eq: Gaussian ML factor model}).\vspace{-8bp}

\begin{lemma}
\label{lem: reformulate gFA} (\cite{khamaru2018computation}) Denoting
$\mathbf{\Phi}=\mathbf{\Psi}^{-1}$, the problem (\ref{eq: Gaussian ML factor model})
is equivalent to minimizing $f\left(\boldsymbol{\phi}\right)$, where
$f\left(\boldsymbol{\phi}\right)=f_{1}\left(\boldsymbol{\phi}\right)-f_{2}\left(\boldsymbol{\phi}\right)$
with $f_{1}\left(\boldsymbol{\phi}\right)=\sum_{i=1}^{p}\left(-\log\phi_{i}+S_{ii}\phi_{i}\right)$,
$f_{2}\left(\boldsymbol{\phi}\right)=-\sum_{i=1}^{r}\left(\log\left(\max\left\{ 1,\lambda_{i}^{*}\right\} \right)-\max\left\{ 1,\lambda_{i}^{*}\right\} +1\right)$
and $\left\{ \lambda_{i}^{*}\right\} _{i=1}^{r}$ are the top $r$
eigenvalues of $\mathbf{S}^{*}=\mathbf{\Phi}^{\frac{1}{2}}\mathbf{S}\mathbf{\Phi}^{\frac{1}{2}}$.
Besides, $f_{1}\left(\boldsymbol{\phi}\right)$ and $f_{2}\left(\boldsymbol{\phi}\right)$
are both convex in $\boldsymbol{\phi}$.
\end{lemma}\vspace{-8bp}

\noindent By linearizing the $f_{2}\left(\boldsymbol{\phi}\right)$
using its sub-gradient, we can majorize $f\left(\boldsymbol{\phi}\right)$
by $\bar{f}\left(\boldsymbol{\phi}\right)=\sum_{i=1}^{p}\big(-\log\phi_{i}+S_{ii}\phi_{i}-\triangledown_{i}^{\left(k\right)}\phi_{i}\big)$,
where $\boldsymbol{\triangledown}^{\left(k\right)}$ is a subgradient
of $f_{2}\left(\boldsymbol{\phi}\right)$ at the $k$th iteration.
The $\boldsymbol{\triangledown}^{\left(k\right)}$ can be calculated
as $\triangledown_{i}^{\left(k\right)}=\big(\mathbf{\Phi}^{-\frac{1}{2}}\mathbf{U}^{*}\mathbf{D}_{1}\mathbf{U}^{*T}\mathbf{\Phi}^{\frac{1}{2}}\mathbf{S}\big)_{ii}$
where $\mathbf{U}^{*}\mathbf{\Lambda}^{*}\mathbf{U}^{*T}$ is the
EVD of $\mathbf{\Phi}^{\frac{1}{2}}\mathbf{S}^{\left(k\right)}\mathbf{\Phi}^{\frac{1}{2}}$
and $\mathbf{D}_{1}=\textrm{Diag}\left(d_{1},\dots,d_{r},0,\dots0\right)$
with $d_{i}=\max\left\{ 0,1-1/\lambda_{i}^{*}\right\} $ \cite{khamaru2018computation,lewis1996derivatives}.
Then the update of $\mathbf{\Phi}$ can be easily obtained as $\phi_{i}^{\left(k+1\right)}=\big(S_{ii}^{\left(k\right)}-\triangledown_{i}^{\left(k\right)}\big)^{-1}$
for $i=1,\dots,p$. By iteratively taking the above procedure, we
can get a converged sequence of $\boldsymbol{\phi}^{\left(k\right)}$.
Finally we can set $\mathbf{\Psi}=\mathbf{\Phi}^{-1}$ and compute
$\mathbf{B}$ via Lemma \ref{lem: optimal B given Psi}. 

It should be noted that the two algorithms are the same and can be
verified by matrix algebra.\vspace{-12bp}

\section{Problem Statement\vspace{-8bp}
}

The $p$-dimensional multivariate Student's $t$ distribution, denoted
as $\boldsymbol{t}_{p}\left(\boldsymbol{\mu},\mathbf{\Sigma},\nu\right)$,
has the probability density function (pdf)\vspace{-8bp}

\begin{equation}
f\left(\mathbf{x}|\boldsymbol{\theta}\right)=\frac{\mathrm{\mathrm{\Gamma}}\left(\frac{\nu+p}{2}\right)}{\mathrm{\mathrm{\Gamma}}\left(\frac{\nu}{2}\right)\nu^{\frac{p}{2}}\pi^{\frac{p}{2}}\lvert\mathbf{\Sigma}\rvert^{\frac{1}{2}}}\left[1+\frac{1}{\nu}\left(\mathbf{x}-\boldsymbol{\mu}\right)^{T}\mathbf{\Sigma}^{-1}\left(\mathbf{x}-\boldsymbol{\mu}\right)\right]^{-\frac{\nu+p}{2}}\label{eq: MVT pdf}
\end{equation}
where $\boldsymbol{\theta}=\left(\boldsymbol{\mu},\mathbf{\Sigma},\nu\right)$,
$\nu$ is the degrees of freedom, $\mathbf{\Sigma}$ is the scale
$p\times p$ positive definite matrix, $\boldsymbol{\mu}$ is the
$p$-dimensional mean vector, and $\mathrm{\mathrm{\Gamma}}\left(a\right)=\int_{0}^{\infty}t^{\left(a-1\right)}\exp\left(-t\right)dt$
is the gamma function. Note that the covariance matrix of $\mathbf{x}$
is $\frac{\nu}{\nu-2}\mathbf{\Sigma}$, and it is not defined for
$\nu\le2$. Interestingly, the above multivariate Student's $t$ distribution
can be represented in a hierarchical structure as $\mathbf{x}|\tau\overset{i.i.d}{\sim}\mathcal{N}_{p}\left(\boldsymbol{\mu},\frac{1}{\tau}\mathbf{\Sigma}\right)$
with $\tau\overset{i.i.d}{\sim}\textrm{Gamma}\left(\frac{\nu}{2},\frac{\nu}{2}\right)$,
where $\mathcal{N}_{p}\left(\boldsymbol{\mu},\mathbf{\Sigma}\right)$
is the multivariate Gaussian distribution with mean vector $\boldsymbol{\mu}$
and covariance matrix $\mathbf{\Sigma}$. $\textrm{Gamma}\left(a,b\right)$
means gamma distribution with shape $a$ and rate $b$, whose pdf
is $f\left(\tau\right)=b^{a}\tau^{\left(a-1\right)}\exp\left(-b\tau\right)/\mathrm{\mathrm{\Gamma}}\left(a\right)$.

We consider that the observed $p$-dimensional data $\mathbf{x}_{t},t=1,\dots,T$
follows the independent and identical distributed (i.i.d.) multivariate
Student's $t$ distribution, i.e., $\mathbf{x}_{t}\sim\boldsymbol{t}_{p}\left(\boldsymbol{\mu},\mathbf{\Sigma},\nu\right)$.
Besides, we assume that $\mathbf{x}_{t}$ follows the FA model, which
means $\mathbf{\Sigma}=\mathbf{B}\mathbf{B}^{T}+\mathbf{\Psi}$. Note
that we omit a scaling factor in order to simplify the notation (recall
that here $\mathbf{\Sigma}$ refers to the scale matrix).  A natural
approach is to obtain the parameter estimation through MLE method,
i.e., maximizing $L\left(\boldsymbol{\theta}|\mathbf{X}\right)=\sum_{t}\log f\left(\mathbf{x}_{t}|\boldsymbol{\theta}\right)$
w.r.t. $\boldsymbol{\theta}$, where $\mathbf{X}\in\mathbb{R}^{T\times p}$
with $\mathbf{x}_{t}$ along the $t$-th row.\vspace{-8bp}

\section{Problem Solution\vspace{-8bp}
}

It is very difficult to directly solve the above MLE problem as the
objective function and constraints are both non-convex. The expectation
maximization (EM) algorithm is a powerful iterative method to handle
such problem \cite{moon1996expectation}. By incorporating the latent
data $\mathbf{Z}$, EM can be employed to convert the maximization
for $L\left(\mathbf{X}|\boldsymbol{\theta}\right)$ to the maximization
for a sequence of simpler and solvable problems. In each iteration,
it requires $Q(\boldsymbol{\theta}|\boldsymbol{\theta}^{\left(k\right)})$,
which is the expected log-likelihood function of $L\left(\mathbf{X}|\boldsymbol{\theta}\right)$
with respect to the current conditional distribution of $\mathbf{Z}$
given the $\mathbf{X}$ and the current estimate of the parameter
$\boldsymbol{\theta}^{\left(k\right)}$. Then it finds $\boldsymbol{\theta}^{\left(k+1\right)}$
by maximizing $Q(\boldsymbol{\theta}|\boldsymbol{\theta}^{\left(k\right)})$.
However, the computational cost of solving the subproblem might still
be rather heavy and make the whole EM algorithm impractical. That
is when the GEM algorithm can help. The GEM is an iterative method
based on the EM philosophy but requiring an improvement at each iteration
instead of a full maximization as in EM.\vspace{-10bp}

\subsection{The RFA Algorithm\vspace{-4bp}
}

In this section, we propose a robust factor analysis (RFA) algorithm
to solve the above problem. By incorporating the latent variables
$\tau$ from the multivariate Student's $t$ hierarchical structure,
the log-likelihood function of the complete data $\left(\mathbf{X},\boldsymbol{\tau}\right)$
is given in (\ref{eq: neg-log MVT hie struc}). Note that $\boldsymbol{\tau}$
corresponds to $\mathbf{Z}$ in our application.\vspace{-15bp}

\begin{equation}
\begin{aligned}L\left(\boldsymbol{\theta}|\mathbf{x},\boldsymbol{\tau}\right) & =\frac{T}{2}\log\lvert\mathbf{\Sigma}^{-1}\rvert-\frac{1}{2}\textrm{Tr}\left(\mathbf{\Sigma}^{-1}\sum_{t=1}^{T}\tau_{t}\left(\mathbf{x}_{t}-\boldsymbol{\mu}\right)\left(\mathbf{x}_{t}-\boldsymbol{\mu}\right)^{T}\right)\\
 & \quad+\frac{T\nu}{2}\log\frac{\nu}{2}+\frac{\nu}{2}\sum_{t=1}^{T}\left(\log\tau_{t}-\tau_{t}\right)-T\log\mathrm{\mathrm{\Gamma}}\left(\frac{\nu}{2}\right)+const.
\end{aligned}
\label{eq: neg-log MVT hie struc}
\end{equation}
\vspace{-25bp}

\subsubsection{Expectation Step:}

The expectation step of the GEM algorithm is to find the conditional
expectation of $L\left(\boldsymbol{\theta}|\mathbf{X},\boldsymbol{\tau}\right)$
over $\boldsymbol{\tau}$ given the observed $\mathbf{X}$ and the
current estimation of $\boldsymbol{\theta}$, i.e., $\boldsymbol{\theta}^{\left(k\right)}$.
Since the conditional expectation of $\tau_{t}$ and $\log\tau_{t}$
for $t=1,\dots,T$ can be directly calculated as\vspace{-15bp}

\[
\begin{aligned}e_{1,t}^{\left(k\right)} & =\textrm{E}\left(\tau_{t}|\mathbf{x}_{t},\boldsymbol{\theta}^{\left(k\right)}\right)=\frac{\nu^{\left(k\right)}+p}{\nu^{\left(k\right)}+d\left(\mathbf{x}_{t},\boldsymbol{\mu}^{\left(k\right)},\mathbf{\Sigma}^{\left(k\right)}\right)}\\
e_{2,t}^{\left(k\right)} & =\textrm{E}\left(\log\tau_{t}|\mathbf{x}_{t},\boldsymbol{\theta}^{\left(k\right)}\right)=\psi\left(\frac{\nu^{\left(k\right)}+p}{2}\right)-\log\frac{\nu^{\left(k\right)}+d\left(\mathbf{x}_{t},\boldsymbol{\mu}^{\left(k\right)},\mathbf{\Sigma}^{\left(k\right)}\right)}{2}
\end{aligned}
\]
where $d\left(\mathbf{x},\boldsymbol{\mu},\mathbf{\Sigma}\right)=\left(\mathbf{x}-\boldsymbol{\mu}\right)^{T}\left(\mathbf{\Sigma}\right)^{-1}\left(\mathbf{x}-\boldsymbol{\mu}\right)$
is the Mahalanobis distance between $\mathbf{x}_{t}$ and $\boldsymbol{\mu}$
\cite{liu1995ml}, then the expectation of the complete data log-likelihood
(\ref{eq: neg-log MVT hie struc}) is \vspace{-16bp}

\begin{equation}
\begin{aligned}Q\left(\boldsymbol{\theta}|\boldsymbol{\theta}^{\left(k\right)}\right) & =\frac{T}{2}\log\lvert\mathbf{\Sigma}^{-1}\rvert-\frac{1}{2}\textrm{Tr}\left(\mathbf{\Sigma}^{-1}\left(\sum_{t=1}^{T}e_{1,t}^{\left(k\right)}\left(\mathbf{x}_{t}-\boldsymbol{\mu}\right)\left(\mathbf{x}_{t}-\boldsymbol{\mu}\right)^{T}\right)\right)\\
 & \quad+\frac{T\nu}{2}\log\frac{\nu}{2}+\frac{\nu}{2}\sum_{t=1}^{T}\left(e_{2,t}^{\left(k\right)}-e_{1,t}^{\left(k\right)}\right)-T\log\mathrm{\mathrm{\Gamma}}\left(\frac{\nu}{2}\right)+const.
\end{aligned}
\label{eq: major stage 1}
\end{equation}
\vspace{-25bp}

\subsubsection{Maximization Step:}

Here we devide the parameters update into two parts.\vspace{-9bp}

\paragraph{Update of $\boldsymbol{\mu}$ and $\nu$:}

 It is easy to see that $\boldsymbol{\mu}$ and $\nu$ are actually
decoupled in $Q\big(\boldsymbol{\theta}|\boldsymbol{\theta}^{\left(k\right)}\big)$
and thus can be easily obtained by setting their derivative to zero.
Then the update scheme for $\left(\boldsymbol{\mu},\nu\right)$ is\vspace{-6bp}

\begin{equation}
\boldsymbol{\mu}^{\left(k+1\right)}=\sum_{t=1}^{T}e_{1,t}^{\left(k\right)}\mathbf{x}_{t}\Big/\sum_{t=1}^{T}e_{1,t}^{\left(k\right)}\label{eq: update scheme for mu}
\end{equation}
\vspace{-25bp}

\begin{equation}
\nu^{\left(k+1\right)}=\underset{\nu>0}{\textrm{argmax}}\left\{ \frac{T\nu}{2}\log\frac{\nu}{2}+\frac{\nu}{2}\sum_{t=1}^{T}\left(e_{2,t}^{\left(k\right)}-e_{1,t}^{\left(k\right)}\right)-T\log\mathrm{\mathrm{\Gamma}}\left(\frac{\nu}{2}\right)\right\} .\label{eq: update scheme for nu}
\end{equation}
According to Proposition 1 in \cite{liu1995ml}, $\nu^{\left(k+1\right)}$
always exists and can be found by bisection search. Interestingly,
the update for $\boldsymbol{\mu}$ and $\nu$ are independent from
each other and irrelevant to $\mathbf{\Sigma}$. \vspace{-6bp}

\paragraph{Update of $\mathbf{\Sigma}$:}

Fixing $\boldsymbol{\mu}$ and $\nu$, maximizing $Q\big(\boldsymbol{\theta}|\boldsymbol{\theta}^{\left(k\right)}\big)$
w.r.t. $\mathbf{\Sigma}$ is reduced to a GFA problem with $\mathbf{S}^{\left(k\right)}=\frac{1}{T}\sum_{t=1}^{T}e_{1,t}^{\left(k\right)}\left(\mathbf{x}_{t}-\boldsymbol{\mu}^{\left(k+1\right)}\right)\left(\mathbf{x}_{t}-\boldsymbol{\mu}^{\left(k+1\right)}\right)^{T}$,
which can be solved by any of the two iterative methods described
in Section \ref{sec: Gaussian-FA-Problem}, which require several
iterations until convergence. Considering that solving $\mathbf{\Sigma}$
exactly would require several iterations and could be time-consuming,
we can instead only run the algorithms for one round, which would
correspond to implementing the GEM instead of EM.\vspace{-4bp}

\subsection{An Acceleration Scheme: PX-EM \vspace{-4bp}
}

A drawback of the EM algorithm is the slow convergence. The parameter
expanded EM (PX-EM) \cite{liu1998parameter} was proposed as an efficient
method to accelerate the classical EM method and can be applied here.
A well-known application of PX-EM on multivariate Student's $t$ case
is to assume that we have $\tau\overset{i.i.d}{\sim}\alpha\textrm{Gamma}\left(\frac{\nu}{2},\frac{\nu}{2}\right)$
in the Student's $t$ hierarchical structure, where $\alpha$ is the
expanded parameter. \vspace{-12bp}

\subsubsection{PX-E}

step: The PX-E step needs only a few modifications on the original
expectation step. Specifically, the Mahalanobis distance should be
calculated as $d\left(\mathbf{x}_{t},\boldsymbol{\mu}_{*},\mathbf{\Sigma}_{*}/\alpha\right)$,
where $\left(x\right)_{*}$ is the corresponding notation of $x$
in the parameter expanded statistical model.\vspace{-12bp}

\subsubsection{PX-M}

step: The update schemes for parameters is similar to those in EM
with only few changes. The update of $\boldsymbol{\mu}$ and $\nu$
keeps unchanged but the update of $\mathbf{\Sigma}$ depends on $\left(\mathbf{S}^{\left(k\right)}\right)_{*}=\alpha^{\left(k\right)}\mathbf{S}^{\left(k\right)}$.
The update of the new parameter $\alpha$ is $\alpha^{\left(k+1\right)}=\frac{\alpha^{\left(k\right)}}{T}\sum_{t=1}^{T}e_{1,t}^{\left(k\right)}$.
After the algorithm achieves convergence, the real parameters should
be recovered as $\boldsymbol{\mu}=\boldsymbol{\mu}_{*}$, $\mathbf{\Sigma}=\mathbf{\Sigma}_{*}/\alpha$
and $\nu=\nu_{*}$.\vspace{-4bp}

\subsection{Robust Enhancement to Missing Data \vspace{-4bp}
}

Due to measurement problems or transmission/storage errors, the observed
data $\mathbf{x}_{t}$ might contain some missing values, which has
been well studied, cf. \cite{liu1995ml}. It turns out that the missing
values can be regarded as latent data like $\boldsymbol{\tau}$. The
new $Q\left(\boldsymbol{\theta}|\boldsymbol{\theta}^{\left(k\right)}\right)$
at expectation step has the same expression w.r.t. $\boldsymbol{\theta}$
as in (\ref{eq: major stage 1}) \cite{liu1995ml}. Therefore the
maximization step can also be achieved by first updating $\boldsymbol{\mu}$
and $\nu$, and then $\mathbf{\Sigma}$ with FA structure imposed.

\vspace{-8bp}

\section{Numerical Experiments\vspace{-8bp}
}

\subsection{Synthetic Data\vspace{-4bp}
}

We generate synthetic data following the multivariate Student's $t$
distribution. The basic dimension setting is $p=100$ and $r=5$.
The true distribution parameters are chosen as follows: $\nu_{\textrm{true}}$
is set to be $7$, the elements of $\boldsymbol{\mu}_{\textrm{true}}$
are drawn i.i.d. from $\mathcal{N}\left(0,1\right)$, $\mathbf{B}_{\textrm{true}}$
comes from BARRA industry model \cite{heckman1999BARRA} with each
sector has size $20$, diagonal elements of $\mathbf{\Psi}_{\textrm{true}}$
are generated independently from an exponential distribution with
mean $10$. The initial $\boldsymbol{\mu}^{\left(0\right)}$, $\nu^{\left(0\right)}$
are set as the sample mean and $10$, and $\mathbf{\Psi}^{\left(0\right)}$,
$\mathbf{B}^{\left(0\right)}$ are given by the naive PCA method.
The convergence condition is set as $\lvert L^{\left(k+1\right)}-L^{\left(k\right)}\rvert\le10^{-6}\lvert L^{\left(k\right)}\rvert$.\vspace{-6bp}

\subsubsection{Convergence Illustration:}

\begin{figure}
\begin{minipage}[t]{0.45\columnwidth}%
\begin{center}
\includegraphics[scale=0.22]{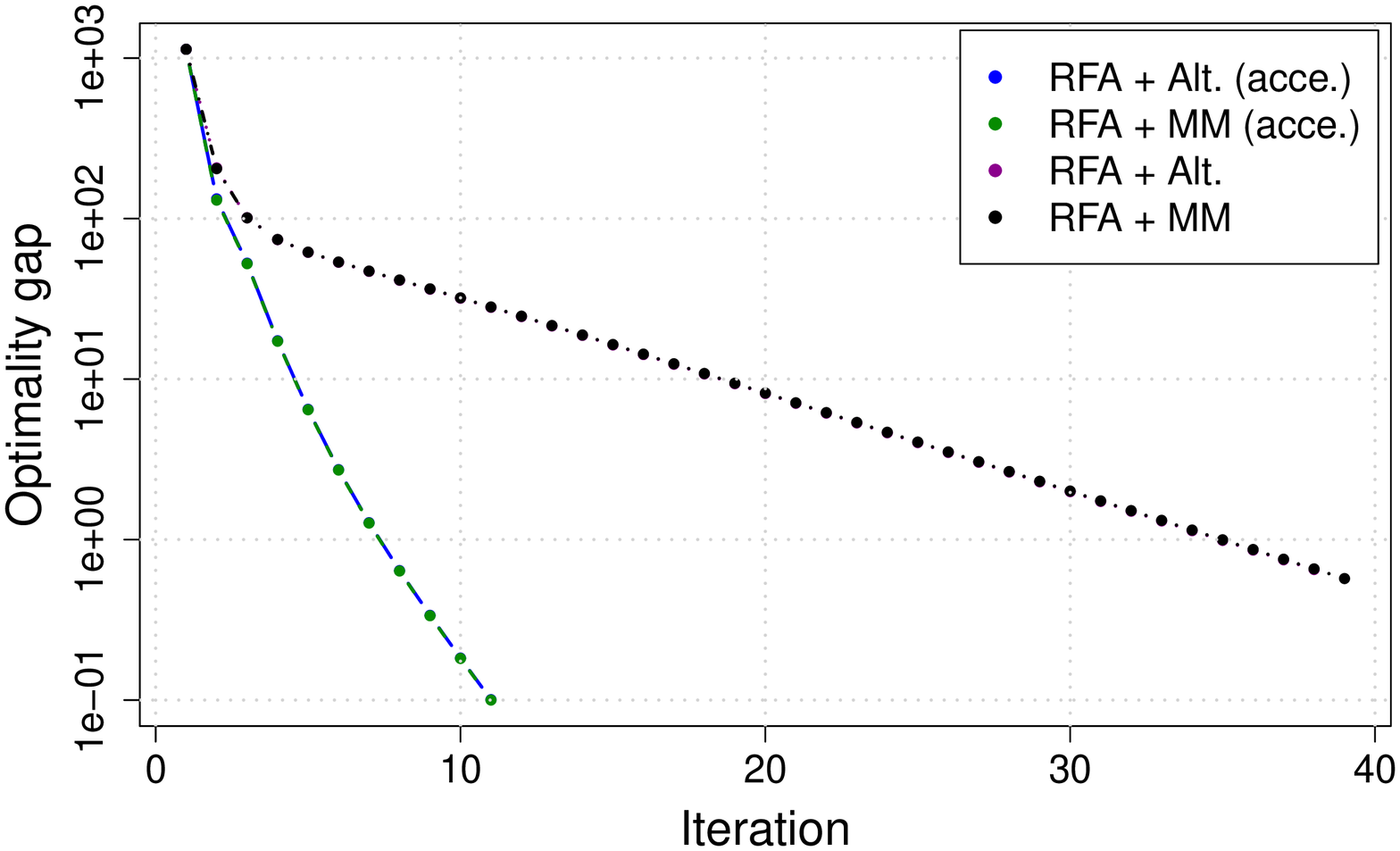}
\par\end{center}
\caption{Optimality gap vs iterations. \label{fig: The-log-likelihood-function}}
\end{minipage}\hfill{}%
\begin{minipage}[t]{0.45\columnwidth}%
\begin{center}
\includegraphics[scale=0.24]{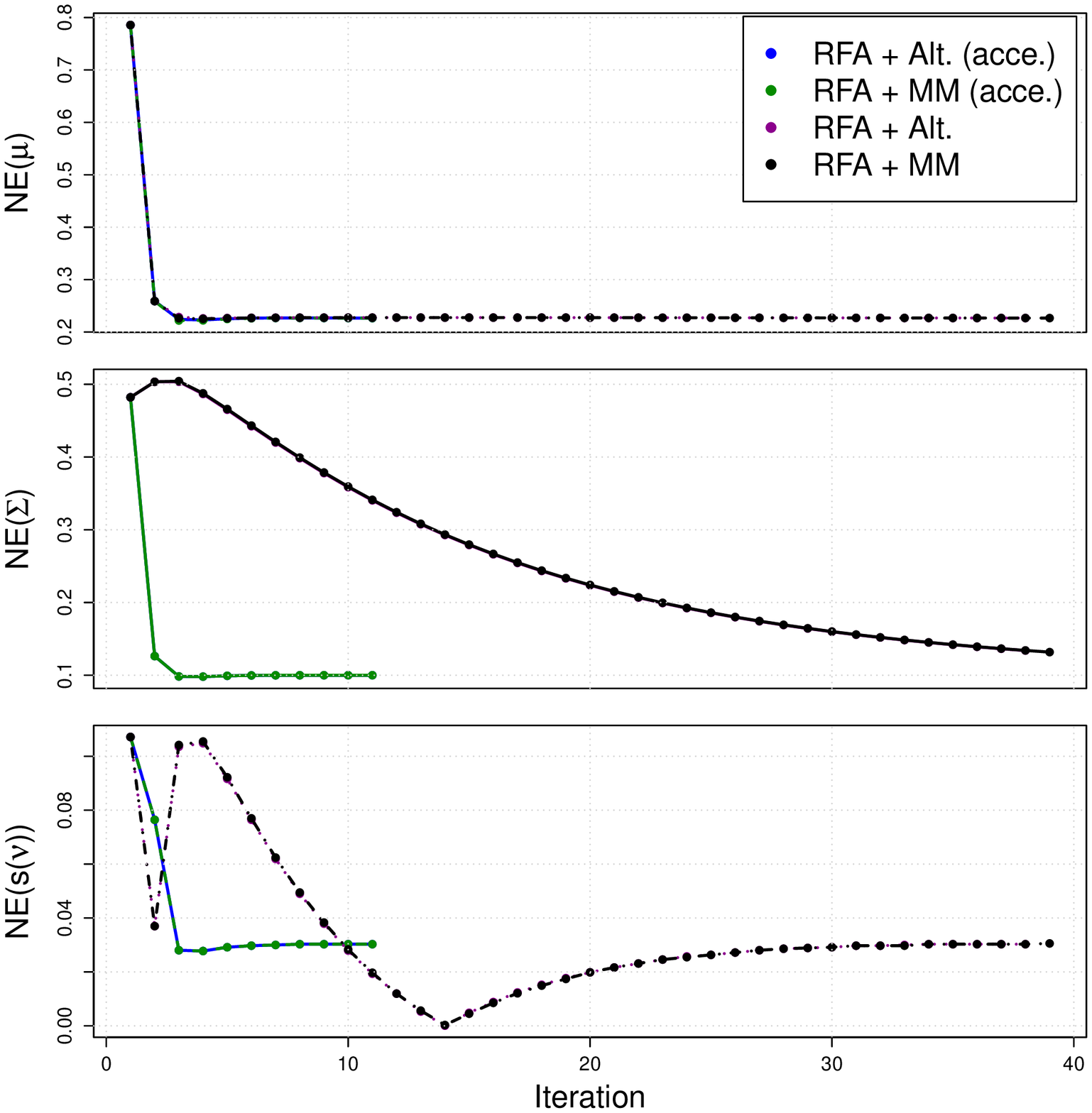}
\par\end{center}
\caption{Estimation error vs iterations. \label{fig: estimation err vs iteration}}
\end{minipage}
\end{figure}

In Fig. \ref{fig: The-log-likelihood-function}, we compare the convergence
of our proposed methods. The log-likelihood function of the observed
data increases monotonically with the iterations and can finally converge.
The PX-EM can significantly accelerate the convergence in this case.
Then, in Fig. \ref{fig: estimation err vs iteration}, we show the
parameter estimation convergence. The parameter estimation normalized
errors are defined as $\textrm{NE}\left(\boldsymbol{\mu}\right)=\lVert\hat{\boldsymbol{\mu}}-\boldsymbol{\mu}_{\textrm{true}}\rVert_{2}/\lVert\boldsymbol{\mu}_{\textrm{true}}\rVert_{2}$,
$\textrm{NE}\left(\mathbf{\Sigma}\right)=\lVert\hat{\mathbf{\Sigma}}-\mathbf{\Sigma}_{\textrm{true}}\rVert_{F}/\lVert\mathbf{\Sigma}_{\textrm{true}}\rVert_{F}$
and $\textrm{NE}\left(\nu\right)=\lvert s\left(\hat{\nu}\right)-s\left(\nu_{\textrm{true}}\right)\rvert/\lvert s\left(\nu_{\textrm{true}}\right)\rvert$
for $\boldsymbol{\mu}$, $\mathbf{\Sigma}$ and $\nu$, where $s\left(\nu\right)=\frac{\nu}{\nu-2}$.
In Figure \ref{fig: estimation err vs iteration}, we can find all
the errors are decreasing and finally converge.

\subsubsection{Robustness Illustration:}

\begin{figure}
\begin{minipage}[t]{0.3\columnwidth}%
\begin{center}
\includegraphics[scale=0.18]{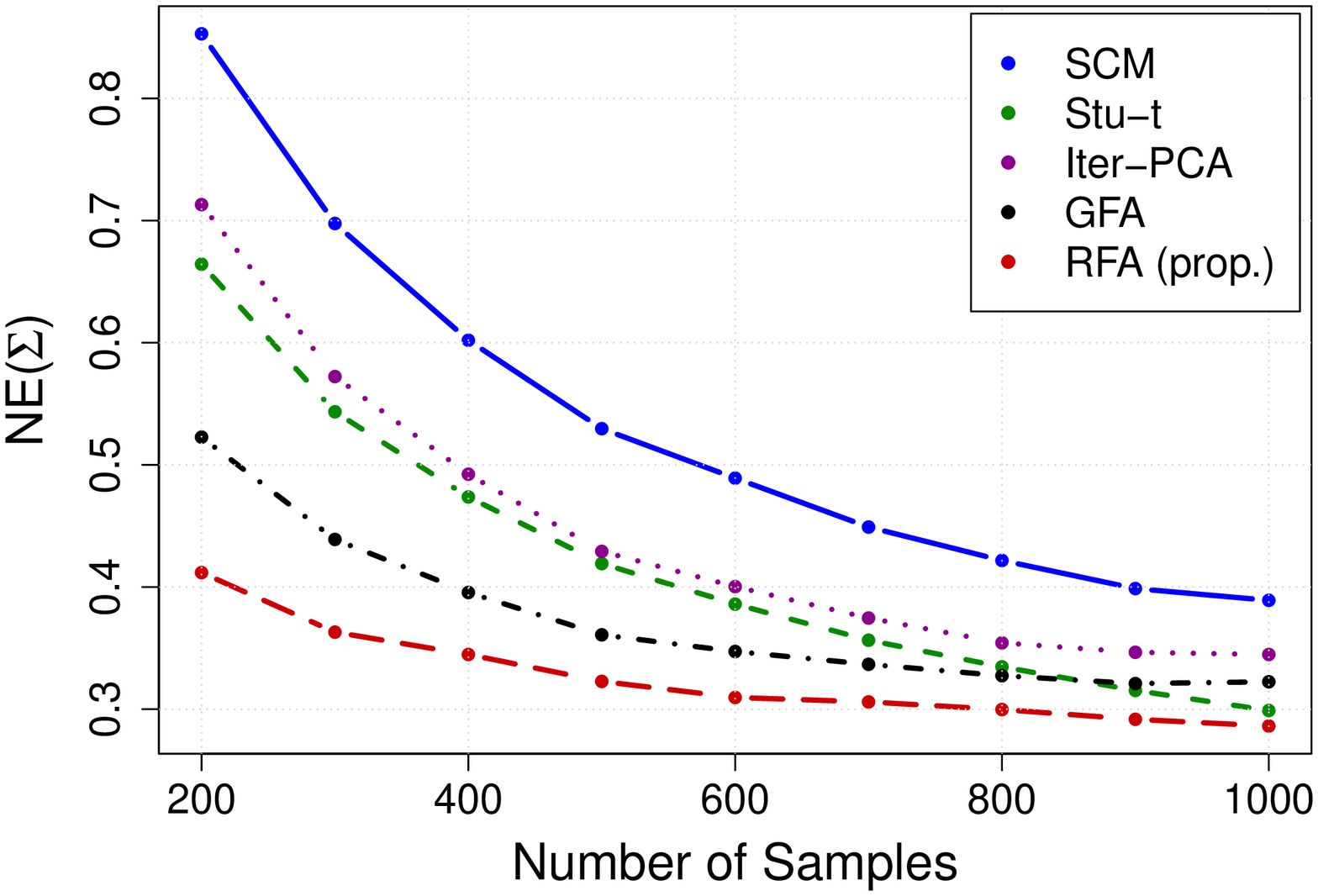}\vspace{-15bp}
\par\end{center}
\caption{Average $\textrm{NE}\left(\mathbf{\Sigma}\right)$ when $n$ changes.
\label{fig: ave error vs npratio}}
\end{minipage}\hfill{}%
\begin{minipage}[t]{0.3\columnwidth}%
\begin{center}
\includegraphics[scale=0.18]{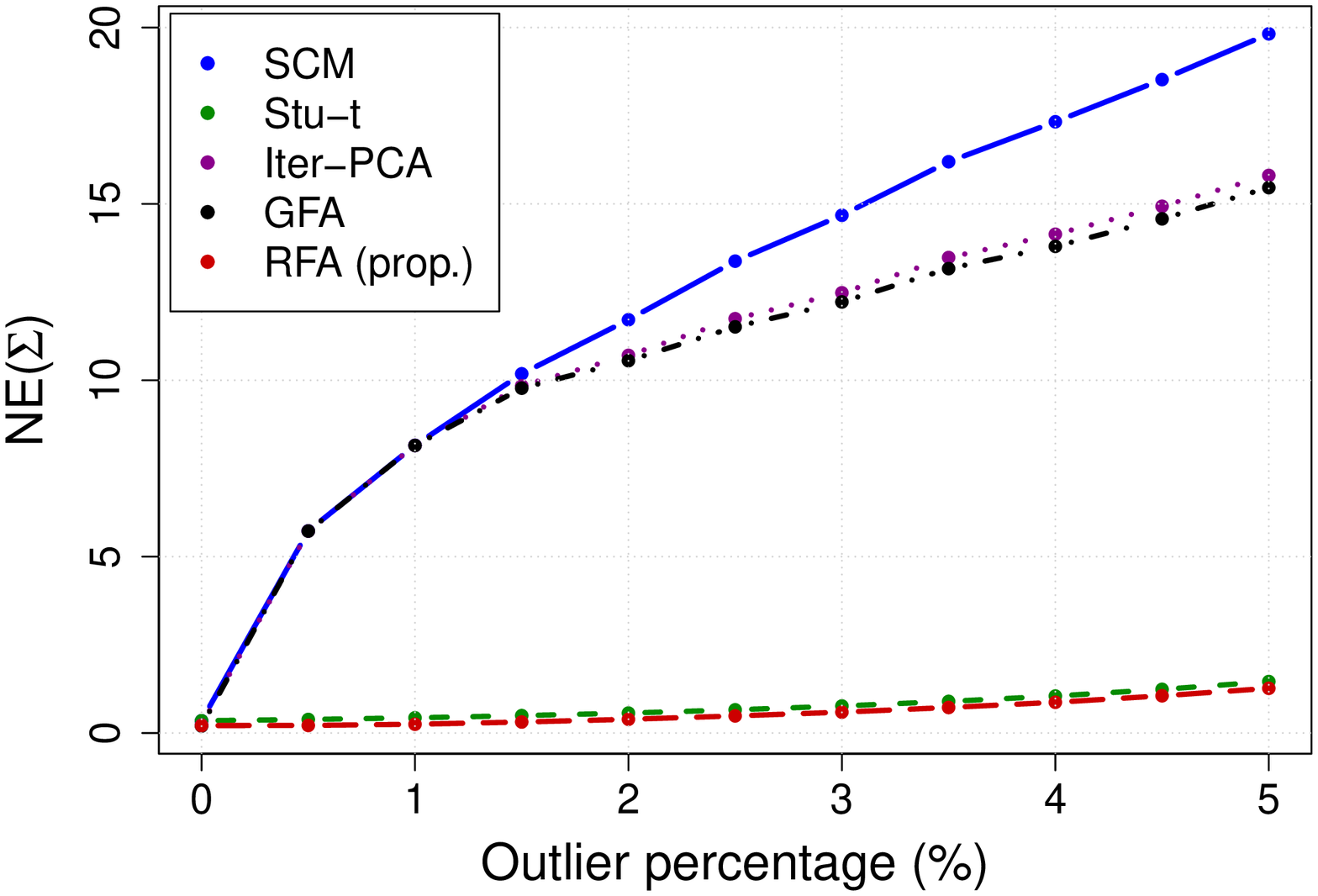}\vspace{-15bp}
\par\end{center}
\caption{Average $\textrm{NE}\left(\mathbf{\Sigma}\right)$ with outliers.\label{fig: estimation err vs outliers ratio}}
\end{minipage}\hfill{}%
\begin{minipage}[t]{0.3\columnwidth}%
\begin{center}
\includegraphics[scale=0.18]{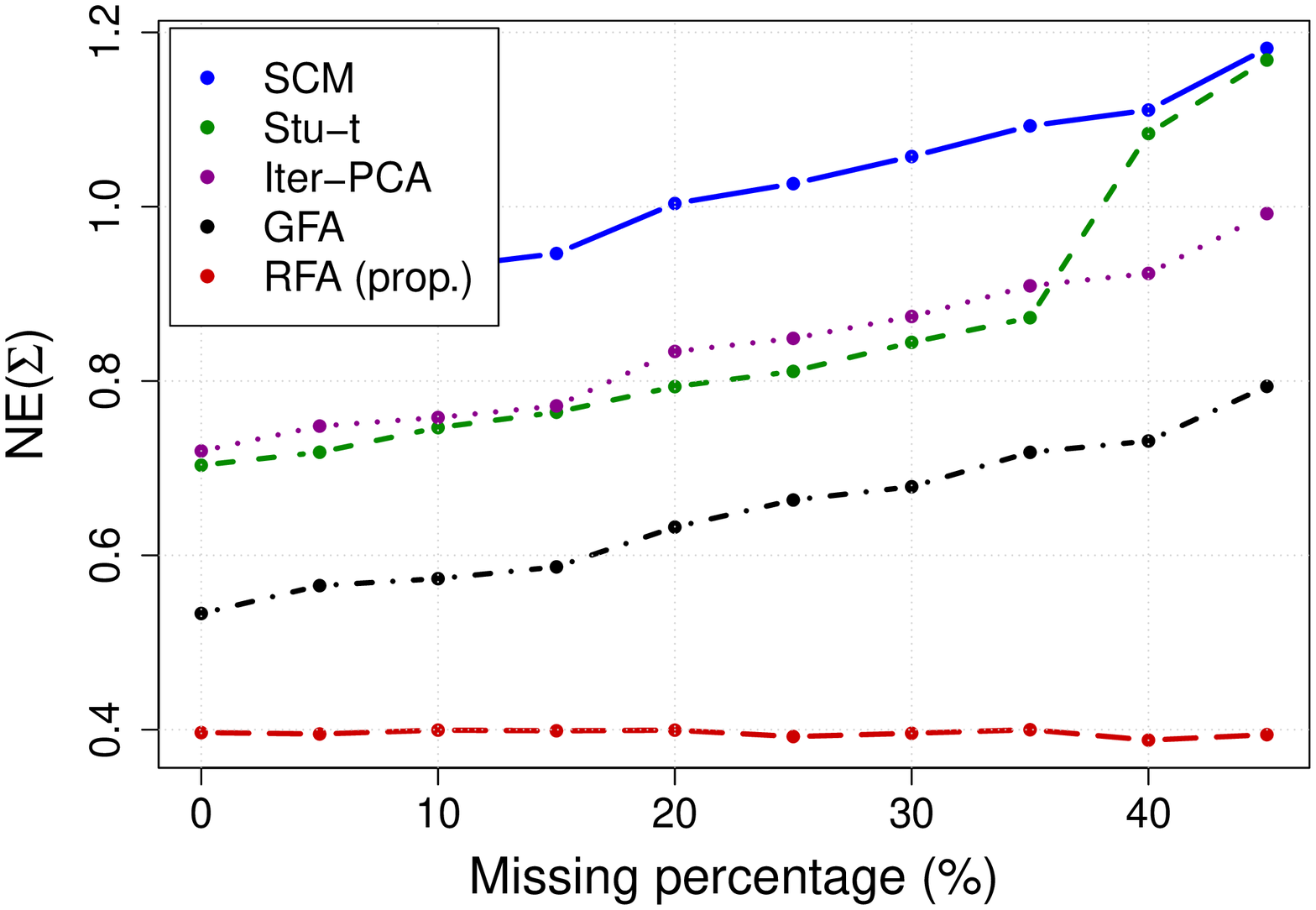}\vspace{-15bp}
\par\end{center}
\caption{Average $\textrm{NE}\left(\mathbf{\Sigma}\right)$ with missing values.\label{fig: estimation err vs missing ratio}}
\end{minipage}

\end{figure}

We compare our proposed RFA methods in covariance matrix estimation
with sample covariance matrix (SCM), Student's $t$ (Stu-t) estimation
(without FA structure) \cite{liu1995ml}, GFA, and iterative PCA (Iter-PCA)
estimation. The results shown below are averaged over $100$ different
realizations of $\mathbf{X}$ following Gaussian distribution. In
Fig. \ref{fig: ave error vs npratio}, we change the sample number
$n$ but fixing $p=100$. All methods show better performance when
$n$ goes large while our proposed RFA method always gives the best
result. In Fig. \ref{fig: estimation err vs outliers ratio}, we randomly
pick some rows of $\mathbf{X}$ and element-wisely add outliers drawn
from $\mathcal{N}\left(0,50\right)$. It is significant that our proposed
RFA method can still hold a good estimation while the results from
non-robust estimation methods are totally destroyed. The results owe
to the robustness of Student's $t$ assumption in resisting the outliers.
In Fig. \ref{fig: estimation err vs missing ratio}, we randomly pick
some rows of $\mathbf{X}$ and randomly set $10\%$ values be missing
for each row. Our proposed robust FA algorithm will be directly fed
with incomplete data while for other methods we need to manually remove
the rows containing the missing values. It is impressive that our
proposed robust FA method can keep the lowest and almost unchanged
performance.

\vspace{-15bp}

\vspace{-28bp}

\subsection{Real Data\vspace{-4bp}
}

\begin{figure}
\begin{minipage}[t]{0.45\columnwidth}%
\begin{center}
\includegraphics[scale=0.25]{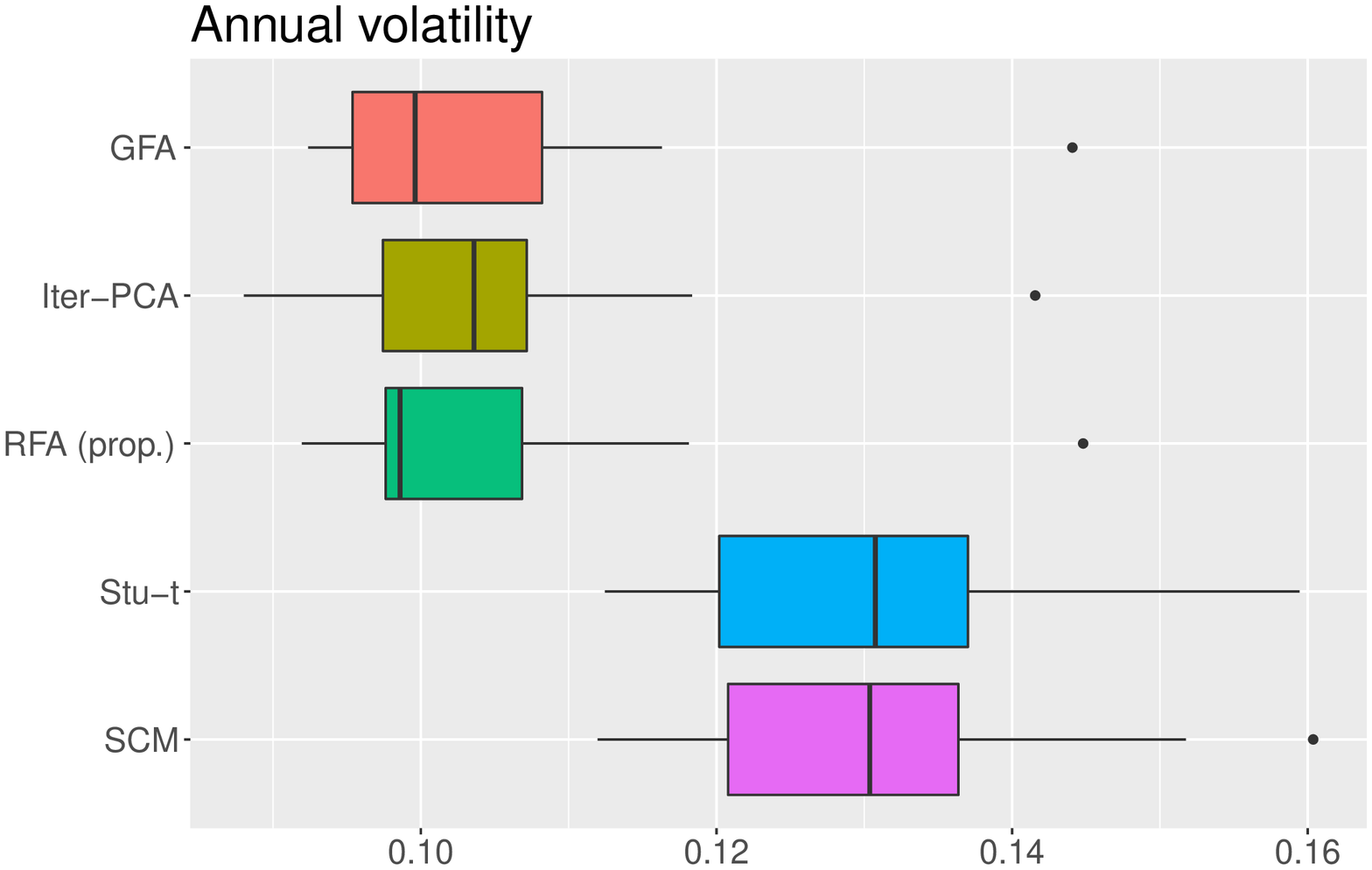}\vspace{-15bp}
\par\end{center}
\caption{Portfolio risk ($r=2$). \label{fig: annual volatility of real data r =00003D 2}}
\end{minipage}\hfill{}%
\begin{minipage}[t]{0.45\columnwidth}%
\begin{center}
\includegraphics[scale=0.24]{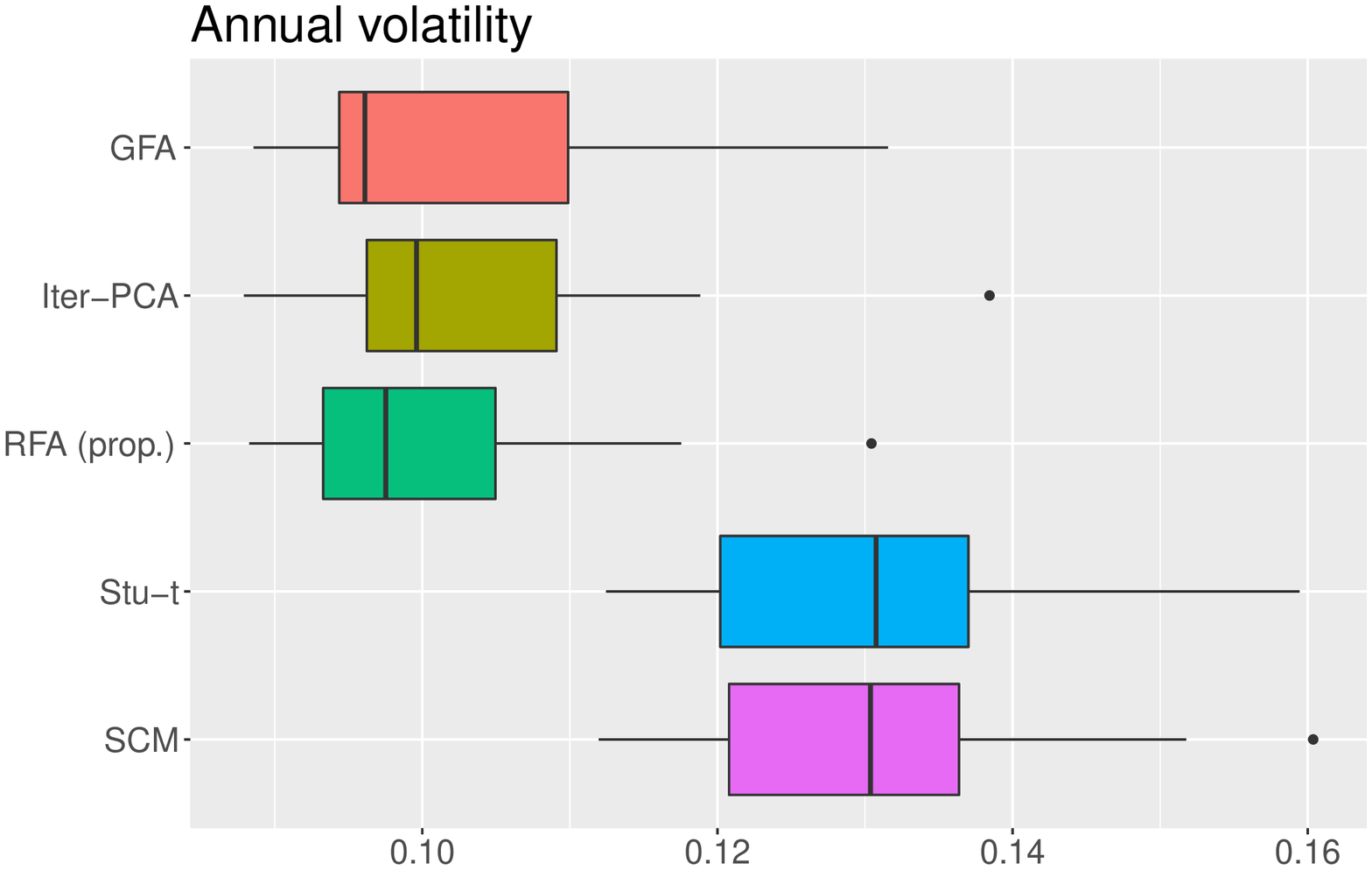}\vspace{-15bp}
\par\end{center}
\caption{Portfolio risk ($r=4$). \label{fig: annual volatility of real data r =00003D 4}}
\end{minipage}
\end{figure}

In this Section, we show the performance of our proposed algorithm
based on the backtest with real financial data. We randomly choose
$50$ stocks for $10$ times from Standard \& Poor's 500 list and
2 years ($2\times252$) continuous historical daily prices data between
01 Dec 2008 and 01 Dec 2018.  Then for each resampling dataset, we
perform the rolling window backtest with lookback window length set
to $100$ days and test window be $5$ days. The rebalance is assumed
to be done everyday without transaction cost. To fairly compare the
estimation performance, we are particularly interested in the global
minimum variance portfolio (GMVP): minimize $\mathbf{w}^{T}\mathbf{\Sigma}\mathbf{w}$
with constraint $\mathbf{1}^{T}\mathbf{w}=1$, where $\mathbf{\Sigma}$
is the covariance matrix obtained from various methods. We respectively
set $r=2$ and $r=4$ in Fig. \ref{fig: annual volatility of real data r =00003D 2}
and \ref{fig: annual volatility of real data r =00003D 4}. It turns
out that our proposed RFA method can achieve smaller out-of-sample
risk with less uncertainty.

\vspace{-15bp}

\section{Conclusion\vspace{-12bp}
}

In this paper, we have proposed the RFA algorithm to obtain the MLE
of the multivariate Student's $t$ distribution with FA structure
imposed. The algorithm was based on the EM framework and had great
estimation accuracy and robustness to outliers and missing values.
The backtest over real financial data has shown advantages and practical
usefulness of our proposed RFA algorithm.

\bibliographystyle{splncs04}
\bibliography{ref_all}

\end{document}